\documentclass[letterpaper, 10 pt, conference]{ieeeconf}  % Comment this line out if you need a4paper

\IEEEoverridecommandlockouts                              % This command is only needed if 
\usepackage{algorithm}
\usepackage{algorithmic}
\usepackage{cite}
\usepackage{xcolor}
  \usepackage[pdftex]{graphicx}
  \graphicspath{ {Image/} }
  \usepackage{caption}
  \usepackage{amssymb}
  \usepackage{siunitx}
  \usepackage{gensymb}
  \usepackage{placeins}
  \usepackage{dblfloatfix}
  \usepackage{graphicx} 
  \usepackage{subcaption}
  \usepackage{mwe}
  \usepackage{todonotes}
  \usepackage[letterpaper,%
            left=0.75in,right=0.75in,top=0.75in,bottom=0.75in,%
            footskip=.25in]{geometry}
\usepackage{amsmath}
\usepackage{tabularx}

\usepackage{algorithm}
\usepackage{algorithmic}
\usepackage{subcaption}

\usepackage{blkarray, bigstrut}

\usepackage{url}

\begin{document}
% \newgeometry{left=0.75in,right=0.75in,top=1in,bottom=0.75in}
\title{A Kinodynamic Aggressive Trajectory Planner For Narrow Passages}
%%%%%% A Kinodynamic Aggressive Trajectory Planner For Narrow Passages

\author{Yaohui Guo$^{1}$, Zhaolun Su$^{2}$, Dmitry Berenson$^{3}$, Ding Zhao$^{4}$% <-this % stops a space
\thanks{Manuscript created September 14, 2017.(Yaohui Guo and Zhaolun Su are co-first authors.)}
\thanks{$^{1}$Y. Guo is with the Robotic Institute at the University of Michigan, Ann Arbor
{\tt\small yaohuig@umich.edu}}% <-this % stops a space
\thanks{$^{2}$Z. Su is with the Department of Electrical and Computer Engineering at the University of Michigan, Ann Arbor 
{\tt\small zhsu@umich.edu}}%
\thanks{$^{3}$D. Berenson is with the Department of Electrical and Computer Engineering and the Robotic Institute at the University of Michigan, Ann Arbor 
{\tt\small berenson@eecs.umich.edu}}%
\thanks{$^{4}$D. Zhao is with the Department of Mechanical Engineering and Robotic Institute at the University of Michigan, Ann Arbor
({\tt\small corresponding author: zhaoding@umich.edu})}%
}

\maketitle

\begin{abstract}

Planning a path for a nonholonomic robot is a challenging topic in motion planning and it becomes more difficult when the desired path contains narrow passages. This kind of scenario can arise, for instance, when quadcopters search a collapsed building after a natural disaster. Choosing the quadcopter as the target platform, this paper proposes the Kinodynamic Aggressive Trajectory (KAT) motion planning algorithm, which aims to compute aggressive trajectories for narrow passages under nonholonomic constraints. This type of maneuvers is necessary because the dynamics of quadcopters entail that some narrow passages can only be traversed at high speed.

To find the best path, the KAT uses RRT to determine a holonomic path first and then adjusts it to satisfy the nonholonomic constraints. The innovations in this process are: 1) The states of the robot are divided into near-holonomic set and non-holonomic set, which makes the constraints local rather than global; 2) For each of the most confined waypoints in the path, KAT plans forward and backward simultaneously around the waypoint to find a feasible local trajectory traversing the narrow passage. Our approach thus transforms a globally-constrained planning problem into a problem with local constraints, and as a result, the computation becomes tractable.
We evaluate KAT by applying it to a quadcopter flying through two inclined holes that require aggressive maneuvers in a simulated environment. The average computation time to successfully find a solution for passing two 50$^{\circ}$ inclined holes is around 32 seconds.  %\todo{It would be helpful to put some numerical result here; i.e. something about computation time.}

% The global feasible trajectory will be computed based on this local planning. KAT innovates by first searching and planning the most difficult part of the path. 

\end{abstract}

%\begin{IEEEkeywords}
%Aggressive, Quadcopter, Maximum Margin Planning, Reverse-Time Planning 
%\end{IEEEkeywords}
\section{Introduction}

A robot's ability to guide itself is the basis for accomplishing higher level tasks, making motion planning a popular and practical problem in robotics. But due to its high computational complexity~\cite{schwartz1987planning}, it is still a challenging problem. The difficulties in planning a path for an informed robot in a complex environment arise from two principal concerns. First, the existence of narrow passages makes sampling method less efficient. Second, some robots have the nonholonomic property, which makes their attainable region a local submanifold of the workspace and thus, similarly, the probability of reaching the goal is low if the robot samples in the whole workspace. Therefore, planning for a nonholonomic robot in an environment with narrow passages could be difficult. 

Quadcopters are typical examples of such a problem. Because of its outstanding mobility, the quadcopter has been used widely in complex and confined environments for applications such as exploration, inspection, and mapping. For instance, a quadcopter is useful for exploring the inside of a collapsed building in a search-and-rescue scenario, where the environment is usually narrow and complex~\cite{michael2012collaborative}. For example, the robot may need to pass through a tilted window without colliding. Unfortunately, a quadcopter can not maintain a tilted attitude at low speed because all the forces applied by the propellers can not entirely offset the gravitational force. Instead, a quadcopter can achieve an instantaneous tilted state by exploiting its dynamics. Thus, an aggressive maneuver is required. %\todo{please make sure this is true. ZS edit: would it be an over-kill to formally prove it here?} 

\begin{figure}[t!]
    \centering
    \includegraphics[width = 1\columnwidth]{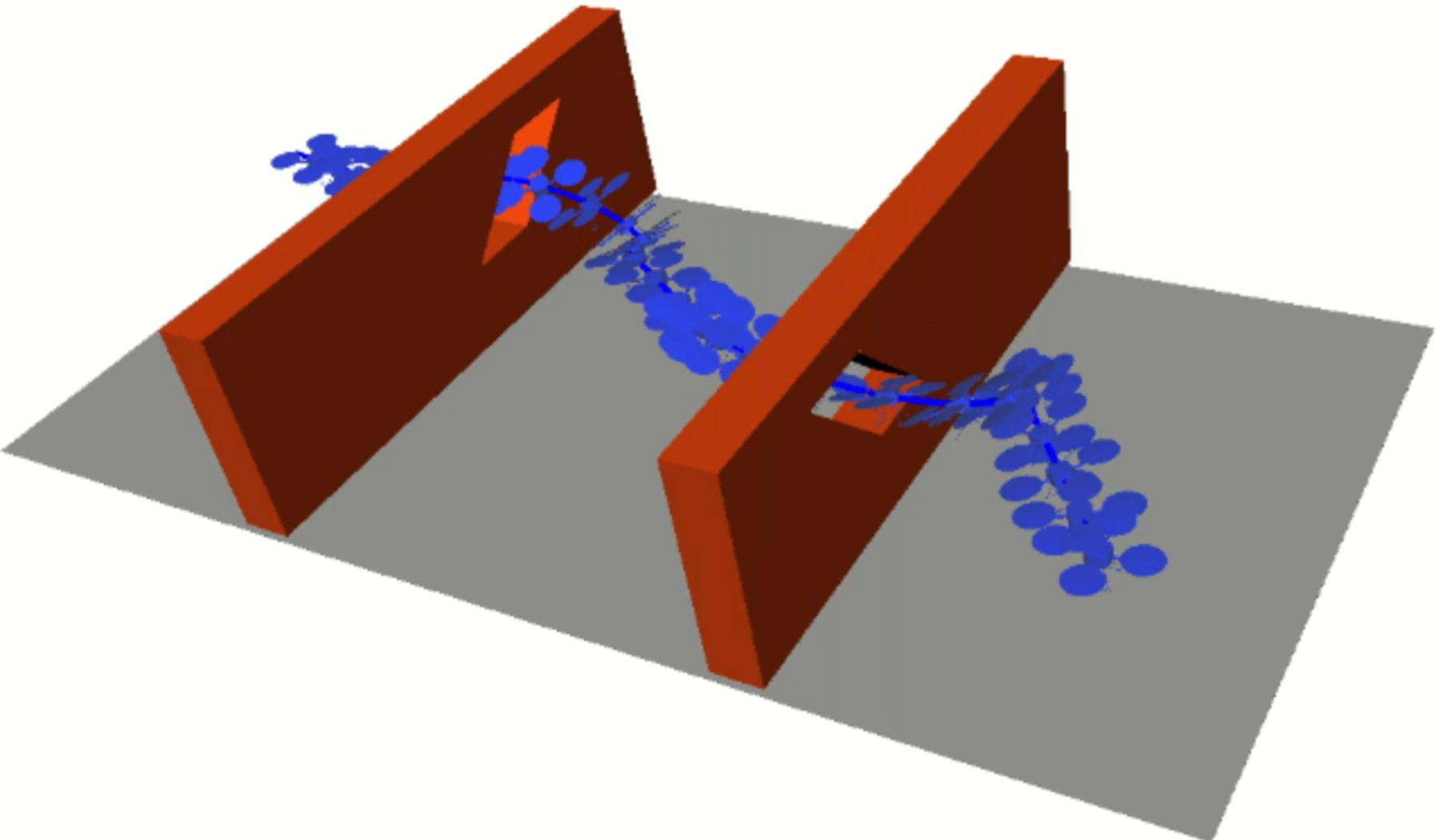}
    \caption{ An aggressive trajectory computed by KAT.}
    \label{fig:wholetra}
\end{figure}

In this paper, we present the Kinodynamic Aggressive Trajectory (KAT) planner for computing a trajectory for a nonholonomic robot in an environment with narrow passages. The main idea of KAT is to eliminate the nonholonomic constraints at the near-holonomic states, and plan the trajectory around narrow points by forward and backward control. This is intended to let the algorithm focus on the bottlenecks of the path, thus the global planning can be broken into several local planning problems. The near-holonomic state assumption assumes that when the quadcopter is flying steadily at very low speed, it is able to change its direction of movement abruptly by a small amount. This assumption allows the quadcopter to move freely in the near-holonomic state and adjust its trajectory to fit the aggressive flying path for the narrow passage.

The paper is organized as follows: section II presents a general overview of related work mainly in the aspect of sampling-based planners; section III describes the math model of the problem; section IV are the details on KAT algorithm; section V presents the experiments we designed for KAT and its results; section VII concludes the work and provides an overview of the future work.

\section{Related Work}

The Rapidly-exploring Random Trees (RRT)~\cite{lavalle2006planning}, as a sampling-based motion planning algorithm, has been widely used for a broad range of robotic systems.
For instance, in~\cite{lavalle2001randomized}, LaValle and Kuffner presented the first randomized approach to systems with kinodynamic constraints. Rather than planning in the configuration space, their approach plans the kinodynamic path in the state space considering the kinodynamic planning as a generalization of holonomic planning. This approach solves the path planning for nonholonomic systems like spacecrafts and hovercrafts, but cannot guarantee optimality and is not efficient for more complex systems due to the high dimensionality of the state space. To improve RRT for finding the optimal path, Karaman and Frazzoli proposed RRT*~\cite{Karaman-RSS-10} for holonomic systems, which grows the same way as RRT except that the tree will locally replan to ensure optimality. In~\cite{karaman2013sampling}, Karaman and Frazzoli proposed an extension of RRT*, which could handle nonholonomic dynamics systems. This algorithm leverages the ball-box theorem to find an optimized extending range for each step while guaranteeing the optimal path. This algorithm works well for nonholonomic planning but is difficult to implement when the system is complex. In~\cite{webb2013kinodynamic}, Webb and Berg introduced the kinodynamic RRT*. Like RRT* algorithm, it is an asymptotically optimal motion planning algorithm, using a fixed-final-state-free-final-time controller to connect any pair of states optimally for systems with controllable linear dynamics to achieve optimality. However, this algorithm still needs to sample the state space, thus it can be time-consuming on systems with high dimensionality. Moon and Chung~\cite{6872545} presented the kinodynamic planner Dual-Tree RRT (DR-RRT) for high-speed navigation of differential drive robot which is composed of a workspace tree and a state tree. The DT-RRT does not reduce the degree of freedom directly. Instead, it searches in the workspace to reduced the search complexity and tries to validate the path in the state space. However, this algorithm is mainly suitable for the low DOF kinodynamic system like Dubin's car and hard to implement on high DOF systems. Other approaches like using motion primitives by building a path set~\cite{knepper2009path}, constructing the state lattice~\cite{pivtoraiko2009differentially}, are to discretize the state space in order to reduce the search complexity. These methods can be used for real-time path planning due to the high efficiency, but compromise the optimality.

Based on the previous work, searching in the state space with the high dimensionality is the bottleneck of most sampling based planners, when dealing with nonholonomic systems. The KAT algorithm we proposed here aims to reduce this complexity by planning a path in the configuration space first and then validating it in the state space.

\section{Problem Statement}
We define our object of interest as a time-invariant dynamic system:
$$ \dot{s}(t) = g(s(t),u(t)),\quad s(0) = s_0$$ 
where $\mathbb{S} \subset \mathbb{R}^{n_s}$ is the state space of the robot; $s(t) = s_t\in \mathbb{S}$ is the state of the robot at time $t$; $\mathbb{U} \subset \mathbb{R}^{n_u}$ is the input space of the system; $u(t)=u_t\in\mathbb{U} $ is the input of the system at time $t$; $g$ is the nonholonomic constraint of the system, which will also be referred as the update function of the system.

For convenience, the following notations are used in this work. The configuration of a robot is the robot's location and attitude; the state of a robot consists of the configuration and the change rate of the configuration. We define the configuration space of the object as $\mathbb{C} \subset \mathbb{R}^{n_c}$. If the robot is free when it is at configuration $c$, we define $c$ as a free configuration; otherwise $c$ is a collision configuration. Define the free configuration space as $\mathbb{C}_{free} \subset \mathbb{C} $ and the collision space as $\mathbb{C}_{col} \subset \mathbb{C} $. Define function $\alpha:\mathbb{S} \rightarrow \mathbb{C}$ maps a state $s$ to its configuration part $c$. Define the free state set as $\mathbb{S}_{free} = \{s \mid \alpha(s) \in \mathbb{C}_{free}\} \subset \mathbb{S}$, and the collision state set as $\mathbb{S}_{col} = \{s \mid \alpha(s) \in \mathbb{C}_{col}\} \subset \mathbb{S}$. 
%\textcolor{red}{define narrow passage}
The narrow configuration set $\mathbb{C}_{nar} \subset \mathbb{C}_{free}$ is the set of all the configurations in the narrow passages. The narrow state set is defined as $\mathbb{S}_{nar} = \{s \mid \alpha(s) \in  \mathbb{C}_{nar} \} \subset \mathbb{S}_{free}$.
Define the start configuration set as $C_s \subset \mathbb{C}_{free}$ and the goal configuration set as $C_g \subset \mathbb{C}_{free}$. Our goal is to find a dynamically feasible path $l:[0,T]\rightarrow S_{free}$ connecting $C_s$ and $C_g$ while passing through some narrow configuration $c_{nar,i},i = 1,2,3,...,n_{nar}$. This is equivalent to $\alpha(l(0)) \in C_s$ and $\alpha(l(T)) \in C_g$, and there exists $t_i\in [0,T], i = 1,2,3,...,n$ such that $\alpha(l(t_i)) = c_{nar,i} \in \{l(t)|t \in [0,T]\},i = 1,2,3,...,n_{nar}$. Also, to satisfy the nonholonomic constraints, there exists valid input $u(t) \in \mathbb{U}$, $t\in [0,T]$ such that
$$l(T)=\int_{0}^{T}g(l(t),u(t))dt + l(0)$$
For numerical computation, we replace the integration with summation, and get
$$l(T)=\sum_{0}^{T}g(s(t),u(t)) \Delta t + l(0)$$

For the quadcopter, we denote its configuration as $c = [p,r]^\intercal $, where $p = [x, y, z]^\intercal\in \mathbb{R}^3$ is the translation of the robot and $r \in SO(3)$ is the rotation. For computational convenience, here we use the quaternion $r= [q_r, q_i, q_j, q_k]^\intercal$ instead of the Euler angles to represent the rotation. Therefore, we have $c= [x, y, z, q_r, q_i, q_j, q_k]^\intercal$. The velocity of a configuration is represented as: $v_c = [ \dot{x}, \dot{y}, \dot{z}, \omega_x, \omega_y, \omega_z]^ \intercal$ . Then a state can be expressed as $s =[p,r,v,\omega]^ \intercal$, where $v = [ \dot{x}, \dot{y}, \dot{z}]^ \intercal$ is the translational velocity and $\omega=[\omega_x, \omega_y, \omega_z]^ \intercal$ is the angular velocity. 
%A testing environment is set to test the algorithm as shown in Fig.\ref{fig:wholetra}, where the quadcopter needs to pass through two walls with inclined holes to reach the goal. The result of the planner should be a series of rotor controls for the quadcopter. 
Because a quadcopter can respond to small changes in its velocity and pose almost instantly by a linear controller when it is still~\cite{brescianini2013quadrocopter}, we can assume the quadcopter is not restricted by the nonholonomic constraints when it is nearly still. Thus here we define the near-holonomic state set as $\mathbb{S}_{holo^*}=\{s|w_{\omega}||\omega||+w_v ||v||+w_r||r-r_0||< \epsilon,s=[p,r,v,\omega]^\intercal \} $ where $w_v$, $w_{\omega}$ and $w_r$ are the weights and $r_0=[1,0,0,0]^\intercal$ is the unit quaternion parallel with \textit{z} axis. The quadcopter with a near-holonomic state means the quadcopter could move freely in any direction within the $ \mathbb{S}_{holo^*}$. In this case, we can constrain the initial and goal states in $\mathbb{S}_{holo^*}$, so the quadcopter does not have to obey nonholonomic constraint when leaving the start configuration and reaching the goal. The nonholonomic constraint is only effective at the states where the quadcopter needs to conduct aggressive maneuver. 

\section{Method} 
The overview of our algorithm appears in Algorithm~\ref{alg:KAT}. The method that we employ consists of four principal parts:
\begin{enumerate}
  \item RRT planning in holonomic space
  \item Sampling narrow configurations with maximum margin in narrow passage
  \item Identifying escape velocity for each narrow configuration
  \item Controller based dual-direction planning with nonholonomic constraints
\end{enumerate}

The goal of the first planning in holonomic space is to efficiently gather information regarding the narrow passage. By testing the robot along the smoothed holonomic path, we could identify the exact location of the narrow passage and collect possible poses that would allow the quadcopter to move through them. 

Since the smoothed trajectory will typically hug the obstacle, it is almost impossible for such a trajectory to be used. We propose implementing the maximum margin sampling inside the narrow passage to avoid such scenarios. From the smoothed holonomic path, we will be able to infer the general configurations where the robot is in a narrow passage. The algorithm will uniformly sample around the cluster centers of the narrow points and replace each cluster center by the configuration with the maximum margin to the surrounding passage. It is obvious that using such a pose is more likely to plan a successful path under nonholonomic constraints. 

Then the algorithm will search a velocity, defined as the escape velocity, to complete the above configuration as a candidate narrow state on the path. To reduce the risk of collision and make it easier for the system to recover to a near-holonomic state, the escape velocity will be the minimum velocity required to pass through the narrow passage. 

Next, starting from the narrow state, a dual-direction controller is employed to find a trajectory through the narrow passage. The dual-direction controller plans both forward and backward the dynamics function. If the planner can reach a near-holonomic point $s_{f} \in \mathbb{S}_{holo^*}$ by forward planning and a near-holonomic point $s_{b} \in \mathbb{S}_{holo^*}$ by backward planning, it will return a local path connecting these two states for the corresponding narrow passage. 

Finally, with the local trajectories through each narrow passage, we can use RRT again to find the paths connecting the start point and end point of all these trajectories sequentially within $\mathbb{S}_{holo^*}$. Thus we have a global path satisfying the nonholonomic constraints. 

 \begin{algorithm}[h]
 \caption{KAT($s_g,s_s, Env,robot$)}
 \label{alg:KAT}
 \begin{algorithmic}[1]
 \renewcommand{\algorithmicrequire}{\textbf{Input:}}
 \renewcommand{\algorithmicensure}{\textbf{Output:}}
 \REQUIRE $C_g,C_s, Env,robot$
 \ENSURE  $ l{global} $
 \\ \textit{Initialization}: KAT $\leftarrow C_g,C_s, Env,robot$
  %\STATE KAT $\leftarrow s_g,s_s, Env$
  \STATE $l_{holo}$ $\leftarrow$ RRT$(C_g,C_s, Env,robot)$
  \STATE  $C_{nar}$ $\leftarrow$ NarrowPoints$(l_{holo},Env,robot)$
  \FOR{ every $c_{nar,i}$ in $C_{nar}$ }
  \STATE $c_{nar,i}$  $\leftarrow$ MaxMargin$(c_{nar,i},Env,robot)$
  \STATE $s_{nar,i}$ $\leftarrow$ $c_{nar,i}$, EscapeVelocity$(c_{nar,i},Env)$
  \STATE  $l_{local,i} \leftarrow$ planFB$(s_{nar,i},Env) $
  \ENDFOR
 \STATE $ l_{global} $ $\leftarrow$ RRTConnectLocalPath$($ all $l_{local,i}) $
 \RETURN $ l_{global} $
 \end{algorithmic}
 \end{algorithm}
 
\subsection{Planning in Holonomic Space with White-listed RRT}
The planner begins by sampling in the holonomic space using the RRT algorithm~\cite{lavalle2006planning}. The purpose of planning in holonomic space is to gain information about the direction and possible poses for crossing the narrow passage. 
    
    \begin{figure}[h]
    \centering
    \includegraphics[width = 0.6\columnwidth]{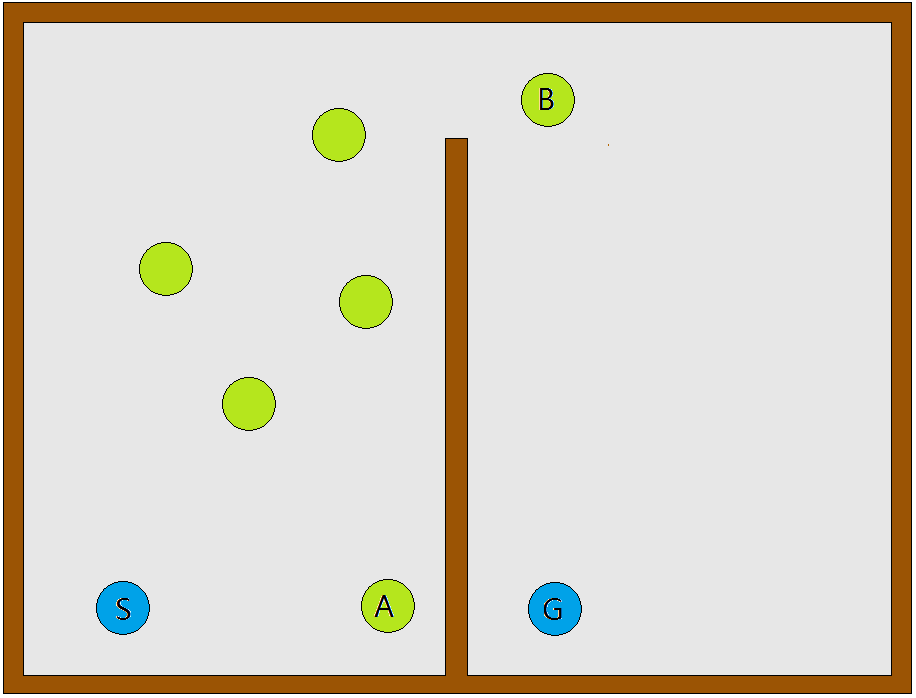}
    \caption{An example showing the advantage of whitelisting. }
    \label{fig:wl}
    \end{figure}
    
In conventional RRT, if the new sample is biased as the goal configuration, the nearest neighbor will be found from the entire explored tree structure. However, we found this algorithm could be inefficient, particularly when a narrow passage presents. For example, in Figure~\ref{fig:wl}, the samples in this 2D environment would not be able to form a direct connection from the goal configuration to the nearest neighbor A. However, it should be able to connect with node B. We eliminate this kind of scenario by adding an additional feature called whitelisting on top of the RRT algorithm. It will keep a list of newly added nodes in the tree structure and make sure every node will only be tested once. Every time RRT samples the goal bias, instead of the entire explored tree structure, the white list will be used to find the nearest neighbor node for testing the goal connectivity and the tested node will be deleted from the it. This means node A in Figure~\ref{fig:wl} will only be tested its connectivity with the goal once and yield for other nodes after it fails. Following this pattern nodes like B will succeed much earlier during planning. %\todo{describe the whitelisting example in the caption of figure 2}

 \begin{algorithm}[h]
 \caption{NarrowPoints($path$,$Env$,$robot$)}
 \label{alg:narrowpoints}
 \begin{algorithmic}[1]
 \renewcommand{\algorithmicrequire}{\textbf{Input:}}
 \renewcommand{\algorithmicensure}{\textbf{Output:}}
 \REQUIRE $path$, $Env$ ,$robot$
 \ENSURE  $C_{nar}$
 \\ \textit{Initialization}: $C_{nar} \leftarrow \emptyset$, $ C'_{nar} \leftarrow \emptyset$
  \FOR { every $c_i$ in $path$ }
  \STATE{set $nbr$ as all 4-connected neighbors of $c_i$}
       \IF {collisionCount$(ngb,Env,robot)>=4$}
    \STATE {$ C'_{nar} \leftarrow  C'_{nar} \cup \{c_i\}$}  
      \ENDIF
 \ENDFOR
  \STATE {$C_{nar}^*$ $\leftarrow$ K-centroids($C'_{nar}$)}
  \STATE {$C_{nar}$ $\leftarrow$ MaxMarginSampling($C_{nar}^*$)}
  \RETURN $C_{nar}$
 \end{algorithmic}
 \end{algorithm}

\subsection{ Maximum Margin Sampling in Narrow Passage }
Since the algorithm plans under delicate conditions, it is preferable to find a way to go through the narrow passage while staying as far from the obstacles as possible. In order to achieve this, each waypoint should be optimized to have the margin to the nearest obstacle maximized. Unfortunately, a smoothed holonomic path would tightly pass through obstacles and leave very little room to work with. The KAT algorithm resolves this problem with a maximum margin sampling scheme. 

After the RRT planning, KAT has found a collision-free holonomic trajectory. The next step is to identify the narrow points on this trajectory. This process is shown in Algorithm~\ref{alg:narrowpoints}. KAT first checks every point on the path and records those that have more than four 4-connected neighbors in $\mathbb{C}_{col}$. Since there may be many points around one narrow passage, KAT uses K-centroids clustering to adaptively select the cluster centers $c_{nar,i}^*$ identified for each narrow passage. The set $C_{nar}^*=\{c_{nar,i}^*\mid i=1,2,3,...,n_{nar} \}$ constitutes the hardest part of the trajectory. 

\begin{figure}
\centering
  \begin{subfigure}{.48\columnwidth}
      \centering
      \includegraphics[width=\columnwidth]{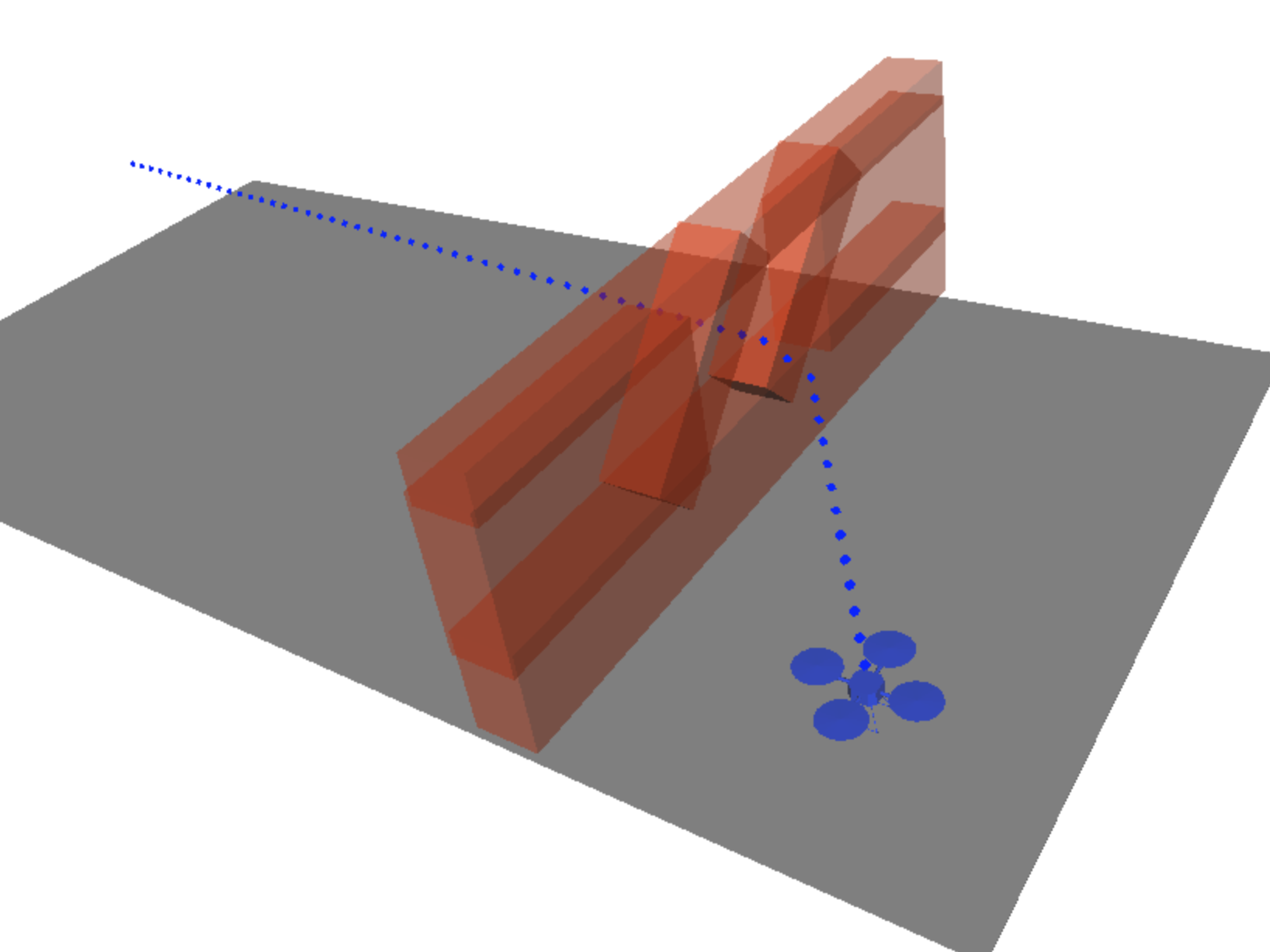}
      \caption{RRT planning in holonomic space}
      \label{fig:rrt}
  \end{subfigure}
  \begin{subfigure}{.48\columnwidth}
      \centering
      \includegraphics[width=\columnwidth]{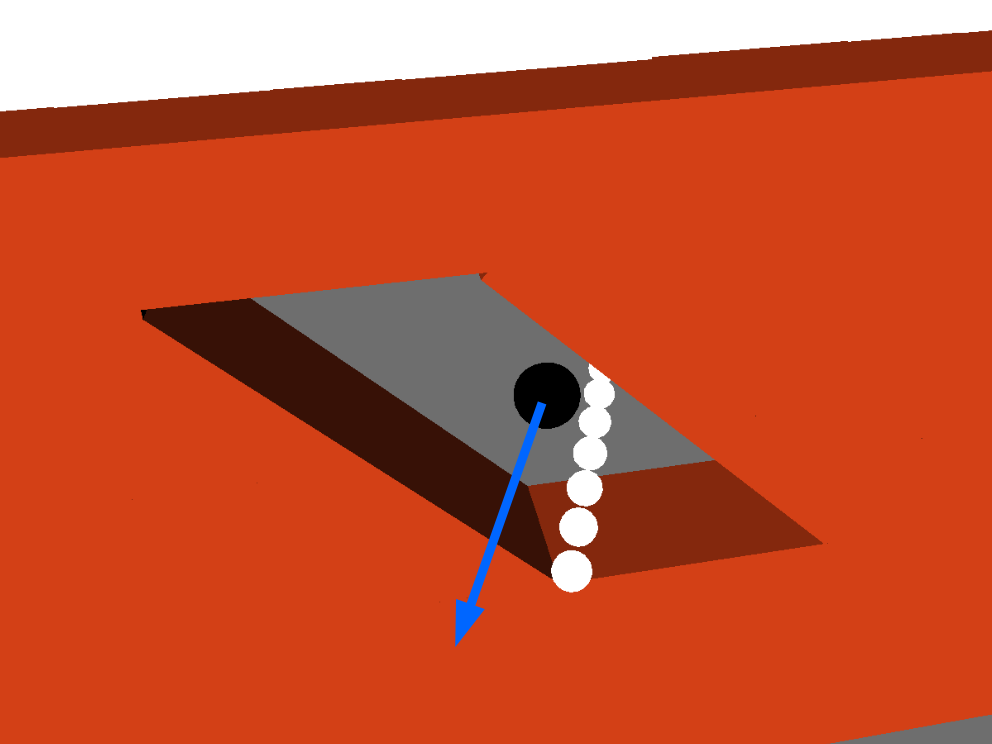}
      \caption{Maximum margin configuration and escape velocity}
      \label{fig:max}
  \end{subfigure}
    \begin{subfigure}{.48\columnwidth}
      \centering
      \includegraphics[width=\columnwidth]{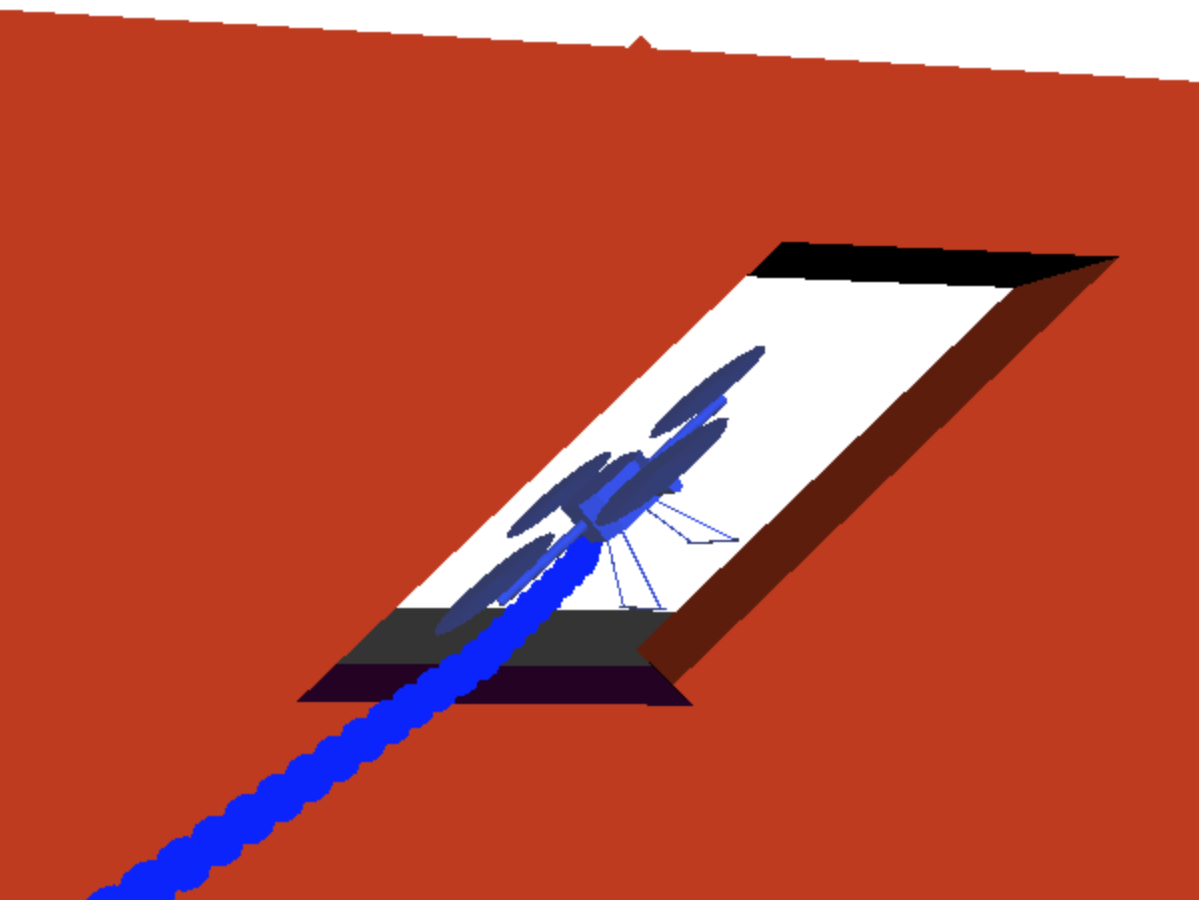}
      \caption{Pose through narrow passage}
      \label{fig:pose}
  \end{subfigure}
    \begin{subfigure}{.48\columnwidth}
      \centering
      \includegraphics[width=\columnwidth]{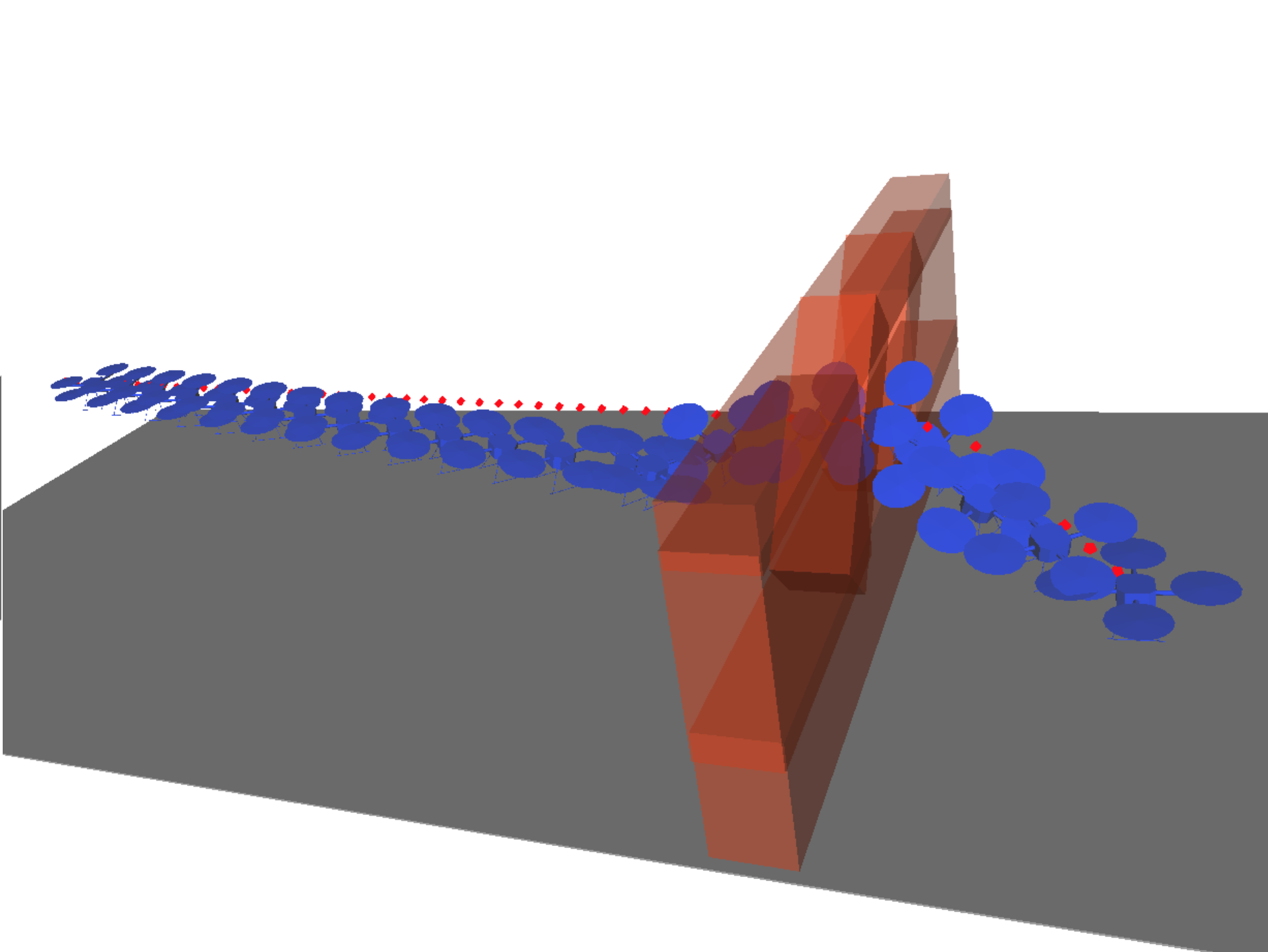}
      \caption{Final trajectory}
      \label{fig:final}
  \end{subfigure}	
  \caption{Four different stages of planning.}
      \label{fig:grouped}
\end{figure}

For each cluster center $c_{nar,i}^*$ generated from Algorithm~\ref{alg:narrowpoints}, KAT will sample configurations uniformly in the plane perpendicular to the planned holonomic path $l_{holo}$, which is denoted as $\mathbb{C}_i^\perp=\{c|c - c_{nar,i}^* \perp l_{holo} \}$. For the sampled poses that are not in collision with an obstacle, we will find the node in them with the lowest objective function value. This objective function calculates the sum of squared distances between this collision-free pose and all of the in-collision poses. The objective function is formulated as:
    \begin{equation*}
    \begin{aligned}
    c_{nar,i} = \mathop{\arg\min}_{c \in \mathbb{C}_{free} \cap \mathbb{C}_i^\perp} \sum_{c_{col} \in \mathbb{C}_{col}}{(c  - c_{col})(c  - c_{col})^T }
    \end{aligned}
    \end{equation*}
Since the value of the objective function is not sensitive to the points far away from the narrow passage, we can simplify the equation above by only considering the collision points near each narrow point. For instance, replace the constraint on $c_{col}$ from $c_{col} \in \mathbb{C}_{col}$ to $c_{col} \in \mathbb{C}_{col} \cap \mathbb{B}_i^\delta$, where $\mathbb{B}_i^\delta =\{c\mid ||c-c_{nar,i}^*||<=\delta\}$ and $\delta$ is a parameter related to the scale of the environment and robot. Fig.\ref{fig:grouped}(b) shows the location of this maximum margin sample derived from the narrow configurations in a passage.

\subsection{Escape Velocity }
In this step, we will complete the narrow point $c_{nar,i}$ in configuration space to a narrow state $s_{nar,i}$ by appending a translation velocity $v_t$ to it while setting the angular velocity to zero. If $v_t$ is the velocity with the minimum norm that can lead the robot through the narrow passage using the forward and backward planning algorithm described in the next part, it is called the escape velocity, denoted by $v_{escape}$. KAT will deduce the direction of $v_{escape}$ by taking the weighted mean of a set of direction vectors, shown in Algorithm~\ref{alg:escapevelocity}. Then the $v_{escape}$ will be fully determined by the forward and backward planning in the next part.

The algorithm first samples a direction set $D = \{d_i|d_i =(x_i,y_i,z_i), |d_i| = 1, i =1,2,3...n\}$ uniformly distributed on the unit sphere $\mathbb{S}^2$. By reusing the cluster center $c_{nar,i}^*$, we can find a heuristic direction $d_{nar,i}$ for each narrow passage. This is to first find the point $c_{nar,i}^{n}$ which is the nearest to $c_{nar,i}^*$ on $l_{holo}$ by $$c_{nar,i}^{n} =  \underset{c\in l_{holo}}{\text{arg\;min}} |c-c_{nar,i}^*|$$ and then identify the tangent direction at $c_{nar,i}^{n}$ by
$$d_{nar,i}=\frac{\text{d}l_{holo}}{\text{d}t_{nar,i}}, \quad l_{holo}(t_{nar,i})  =c_{nar,i}^{n}$$
Next, for each $d_i$ in $D$, if $\langle d_i,d_{nar,i} \rangle<=0$, it will be removed from $D$. After constructing $D$, a length $t_i$ is generated for each $d_i$ following a normal distribution $N(\mu_{nar},\sigma_{nar})$, where $\sigma_{nar}$ and $\mu_{nar}$ should be selected according to the property of the environment, and translate the robot from the narrow point $c_{nar,i}$ by each pair of $(t_i, d_i)$ to $c_{nar,i}^t$. If $c_{nar,i}^t \in C_{col}$, delete $(d_i)$ from $D$. Finally, if $D$ is not empty, the direction of the escape velocity will be calculated as

$$v_{escape,i}^* = \frac{\sum_{d_i \in D}d_it_i}{ \left| {\sum_{d_i \in D}d_it_i} \right| } $$

 \begin{algorithm}[h]
 \caption{EscapeVelocity($c_{nar}$,$Env$,$robot$)}
 \label{alg:escapevelocity}
 \begin{algorithmic}[1]
 \renewcommand{\algorithmicrequire}{\textbf{Input:}}
 \renewcommand{\algorithmicensure}{\textbf{Output:}}
 \REQUIRE $c_{nar}$, $Env$ ,$robot$
 \ENSURE  $v_{escape}^*$
 \\ \textit{Initialization}: $D= \{d_i|d_i\in \mathbb{S}^2, i =1,2,3...n\}$
%  \STATE {generate $D= \{d_i|d_i\in \mathbb{S}^2, i =1...n\}$}
    \STATE{remove all $d_i$ goes against $l_{holo}$ from $D$}
  \FOR{ every $d_i$ in $D$ }
      \STATE {generate $t_i \sim N(\mu_{nar},\sigma_{nar})$}
    \STATE {$c_{nar,i} = translate(c_{nar},d_i,t_i)$}     
       \IF {$c_{nar,i} \in \mathbb{C}_{col}$}
    \STATE {delete $d_i$ from $D$}        
    
      \ENDIF  
 \ENDFOR
 
 \STATE {$v_{escape} = (\sum_{d_i \in D}d_it_i) / \left| {\sum_{d_i \in D}d_it_i} \right| $}
  \RETURN  $v_{escape}^* $
 \end{algorithmic}
 \end{algorithm}

\subsection{Forward and Backward Planning With Controller}

This step finds local dynamically feasible paths through each narrow passage by exploiting $c_{nar,i}$ and $v_{escape,i}^* $ deduced above. If such paths can be found, the bottlenecks of the planning are solved since the remaining task is to connect the starts and ends of each local path to form a global path.

The algorithm here will generate $s_{nar,i}$ by adding $v_{escape,i}$ to $c_{nar,i}$. The direction of $v_{escape,i} $ is determined by $v_{escape,i}^* $ and its norm increases each time. For each generated $s_{nar,i}$ the algorithm will use a forward controller $C_f:(\mathbb{S},\mathbb{S})\rightarrow \mathbb{U}$ and a time-inverse controller $C_b:(\mathbb{S},\mathbb{S})\rightarrow \mathbb{U}$ to stabilize $s_{nar,i}$ respectively. This process should satisfy
    \begin{equation*}
    \begin{aligned}
    &s_{t+1} = g(s_t,u_t)\Delta t + s_t\\
    &u_t=\begin{cases}
               C_f(s_t,s_{still}),\quad t>t_{nar,i}\\
               C_b(s_t,s_{still}),\quad t<=t_{nar,i}
            \end{cases}
    %&s_{t_{nar,i}} = s_{nar,i}\\
    %&t \in [0,T_{nar,i}],\quad t_{nar,i}
    \end{aligned}
    \end{equation*}
for $\forall t \in [0,T_{nar,i}]$, and $\exists t_{nar,i}\in [0,T_{nar,i}]$ such that $s_{t_{nar,i}} = s_{nar,i}$. If $s_0,s_{T_{nar,i}} \in  \mathbb{S}_{holo^*}$ and $s_t\in \mathbb{S}_{free},\forall t \in [0,T_{nar,i}]$, the path $l_{nar,i}(t)=s_t,t\in [0,T_{nar,i}]$ is a local dynamically feasible path through the corresponding narrow passage with both start and end points in $\mathbb{S}_{holo^*}$. If the algorithm can find a local passage $l_{local,i}$ for every $c_{nar,i}$ on $l_{holo}$, the global path $l_{global}$ will be constructed from connecting the start points and end points of these local paths sequentially. Since the start and end of each $l_{local,i}$ are in $\mathbb{S}_{holo^*}$, the algorithm will use RRT to plan the path by sampling in $\mathbb{S}_{holo^*}$ without restricted by the nonholonomic constraints.

\section{Experiments and Results}
This section describes the dynamics model that is used for our experiment, results from different sections of the KAT algorithm as well as the final path generated for different environment settings.

Our algorithms were implemented in Python with Openrave. All experiments were executed on a laptop with an Intel(R) Core(TM) i7-5600U at 2.6GHz, 8GB of RAM. Each experiment ran until a trajectory was found, or 1 minutes had elapsed. We performed 30 trials for each experiment and removed the fastest and slowest. The video of the whole computational processing is shown in~\cite{algovideo1}

\subsection{Dynamic model}
    \begin{figure}[h]
        \centering
        \includegraphics[width = 0.7\columnwidth]{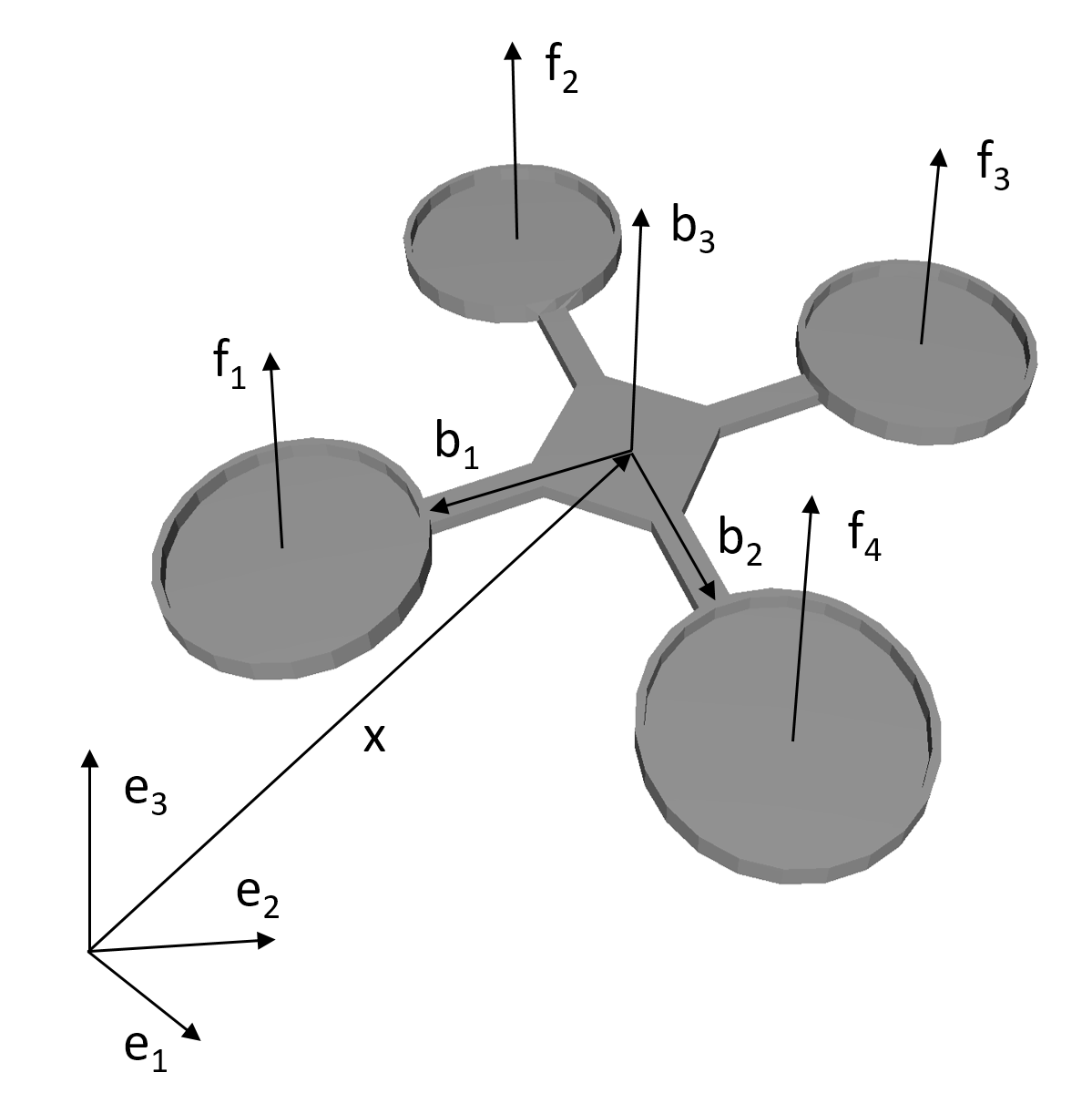}
        \caption{The configuration of the quadcopter.}
        \label{fig:quad}
    \end{figure}
We exploit the quadcopter dynamics model and controller developed in~\cite{loianno2017estimation}. The configuration of the drone model is shown in Figure~\ref{fig:quad}. The inertia frame with axes $e_1$,$e_2$,$e_3$ is a reference frame attached to the ground; the body frame with axes $b_1$,$b_2$,$b_3$ is a frame attached to the drone. The dynamics of the drone is described by the following equations:     
    \begin{equation*}
    \begin{aligned}
        &\dot{x} = v,\quad \dot{v} = mge_3-fRe_3 \\
        &\dot{R} = R\hat{\Omega}, \quad J\dot{\Omega}+\Omega\times J \Omega=M \\
         &\begin{bmatrix}
            f\\M_1\\M_2\\M_3 \end{bmatrix} 
            = \begin{bmatrix}
                1&1&1&1\\
                0 & -d & 0 & d\\ 
                d&0&-d&0\\
                -c&c&-c&c
               \end{bmatrix}
               \begin{bmatrix}
               f_1\\ f_2\\ f_3\\ f_4 
               \end{bmatrix}
    \end{aligned}
    \end{equation*}
where $m\in \mathbb{R}$ is the total mass of the quadcopter; $J\in\mathbb{R}^{3\times3}$ is the inertia matrix with respect to the drone frame; $R\in SO(3)$ is the rotation matrix from the body-fixed frame to the inertial frame; $\Omega\in\mathbb{R}^3$ is the angular velocity in the drone frame; $x\in\mathbb{R}^3$ is the position of the center of mass in the inertial frame; $v\in\mathbb{R}^3$ is the velocity of the center of mass in the inertial frame; $d\in\mathbb{R}$ is the distance from the each axis of the rotor to the drone center; $f_i\in\mathbb{R}$ is the thrust of the $i$th propeller; $f\in\mathbb{R}$ is the total thrust; $\tau_i\in\mathbb{R}$ is the torque applied to the drone by $i$th rotor along $i$th axis; $M\in\mathbb{R}^3$ is the moment vector in the body frame. The controller is modified from~\cite{loianno2017estimation}. The forward controller is

    \begin{equation*}
    \begin{aligned}
        M_f = &-k_\Omega e_\Omega+\Omega \times J \Omega \\
        	%&-J(\hat{\Omega} R_C^\top \Omega_C - R^\top R_C\dot{\Omega}_C)\\
            &+Y(k_{zv}R^\top\hat{\omega}^\top-k_zR)e_3, \\
        f_f = & (-k_ve_v+mge_3) \cdot Re_3    
    \end{aligned}
    \end{equation*}
and the backward controller is 

    \begin{equation*}
    \begin{aligned}
        M_b = &k_\Omega e_\Omega-\Omega \times J \Omega \\
        	%&+J(\hat{\Omega} R_C^\top \Omega_C - R^\top R_C\dot{\Omega}_C) \\
            &-Y(k_{zv}R^\top\hat{\omega}^\top-k_zR)e_3, \\
        f_b = & (k_ve_v+mge_3) \cdot Re_3    
    \end{aligned}
    \end{equation*}
where
    \begin{equation*}
            Y = 
            \begin{bmatrix}
                0&1&0\\
                1&0&0\\ 
                0&0&0
               \end{bmatrix} 
      \end{equation*}
The desired translational velocity and angular velocity are set to zero; $e_\Omega$, $e_v$ are the angular velocity error and translational velocity error; $k_\Omega$, $k_{zv}$, $k_z$, $k_v$ are control parameters;$M_f$, $f_f$, $M_b$, $M_b$ are the total moment and force of forward control and backward control respectively.

\subsection{Experiment Setup}
The experiment setting is shown in Figure~\ref{fig:rrt}. The start configuration and goal configuration are at the different sides of the wall; all paths connecting the two sides contain a narrow passage on the wall, which is an inclined hole and the quadcopter cannot reach the goal without passing it.

\subsection{Experiment Result}
The holonomic path found by the RRT algorithm is shown in Figure~\ref{fig:rrt}. Table~\ref{compholo} shows the comparison of conventional RRT with white-list RRT on the same environment setting. By limiting the computation time to 1 minute, the conventional RRT has a slightly lower success rate than modified RRT. In addition, the white-listed RRT finishes in less time and with fewer sampled nodes.

\begin{table}[h]
\renewcommand{\arraystretch}{1.3}
\caption{Computation time of holonomic sampling}
\label{compholo}
\centering
\begin{tabular}{c||c|c|c|c}
\hline
   & \bfseries Time& \begin{tabular}[x]{@{}c@{}}\bfseries Sampled\\ \bfseries Nodes\end{tabular} & \begin{tabular}[x]{@{}c@{}}\bfseries Success\\\bfseries Rate\end{tabular} & \bfseries Path Length\\
\hline\hline 
 \bfseries RRT & 23.3s & 12197 & 100\% & 9.05\\
 \bfseries Modified RRT  & 17.7s & 9648 & 100\% & 9.03\\
\hline
\end{tabular}
\end{table}

This holonomic path identifies a collision-free trajectory that can connect the goal and start, but violates the dynamics of the quadcopter. One can easily find that the quadcopter should fly at a low speed in order to be able to make a sharp turn near the hole, which conflicts with the need of a high speed pass for the inclined hole. By sampling and clustering adjacent points of each node on the path, the algorithm identifies the narrow passage and refines the configuration for passing, as illustrated in Figure~\ref{fig:max}. From Figure~\ref{fig:max}, we can see the refined point allows the quadcopter to leave a safe margin from the wall. The next step is to sample forward and backward to generate a feasible path passing the narrow passage while making it possible to connect the start and end nodes with a holonomic path. 

\begin{figure}[h]
        \centering
        \includegraphics[width = 0.7\columnwidth]{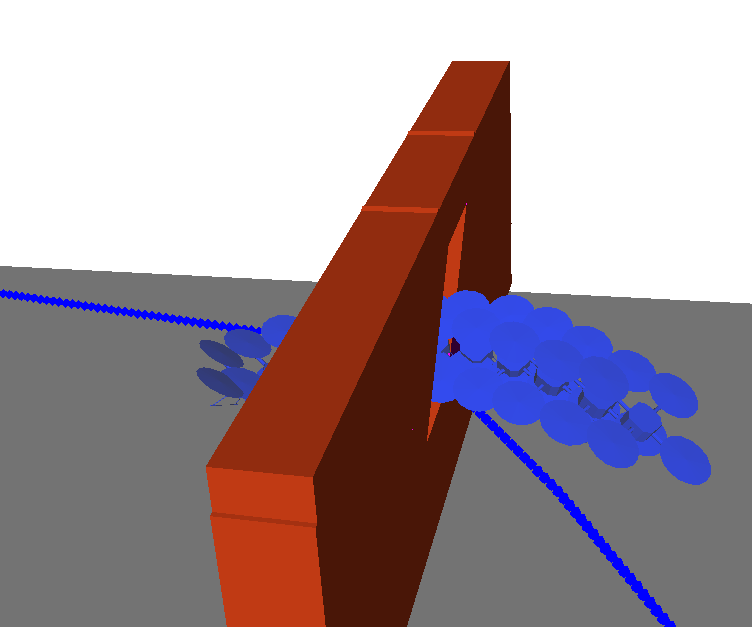}
        \caption{Local path through the narrow passage.}
        \label{fig:local}
    \end{figure}

\begin{figure}[h]
        \centering
        \includegraphics[width = 0.9\columnwidth]{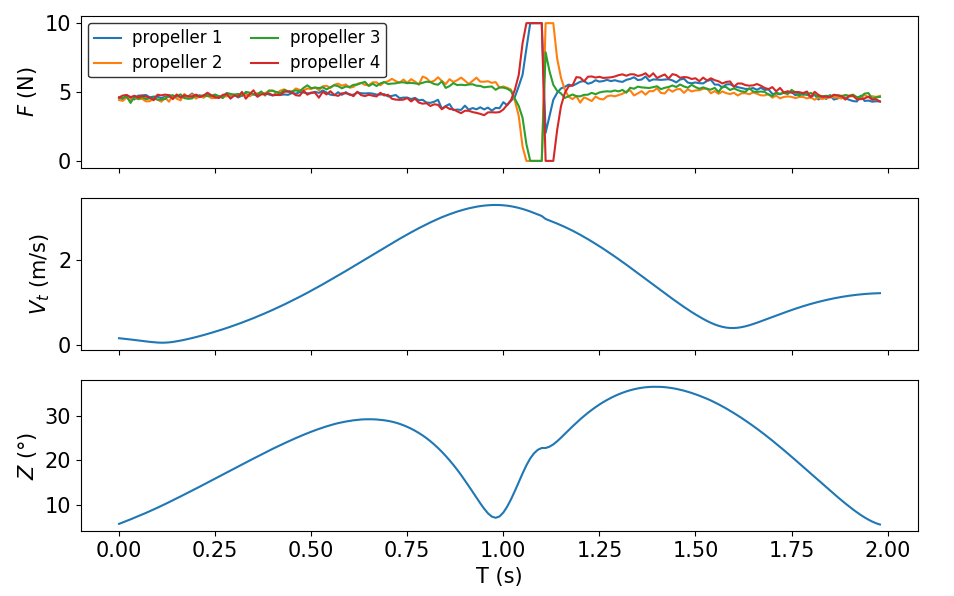}
        \caption{Local path state space analysis.}
        \label{fig:analysis}
    \end{figure}

The local path built here is shown in Figure~\ref{fig:local}. Figure~\ref{fig:analysis} provides an analysis of this process. $F$ is the thrust generated by each propeller; $v_t$ is the translational velocity; $Z$ is the angle between $b_3$ and $e_3$ defined in Figure~\ref{fig:quad}; $t_{nar}$ is the time when the quadcopter passes the narrow passage. The control input saturates when $t = t_{nar}$, because the feedback error reaches its maximum, which is the difference between the instantaneous state and the near-holonomic state. This is very different from planning the path from one side to the other. 

At the last step, KAT connects the end and start points of the local path with the corresponding nearest nodes in the holonomic path and returns the result, as shown in Figure~\ref{fig:final}.  

For testing the effectiveness of KAT, we derived a more complex environment setup. By setting up two obstacle walls, each with a different window opening angle, we proved that KAT could connect multiple aggressive trajectories. Figure~\ref{fig:wholetra} shows the finished trajectory for passing these two obstacles. In~\cite{algovideo1}, the video demonstrates the entire planning process and simulated execution of the planned trajectory. We also changed the opening angle on both walls and analyzed how it would affect the results. Table~\ref{comptable} shows that as the opening gets steeper, the maximum speed that KAT has to sample rises. As the opening becomes near-vertical, the maximum speed of the trajectory will be over 10 meters per second.

\begin{table}[h]
\renewcommand{\arraystretch}{1.3}
\caption{Average result of 30 trials for dual-obstacle setup}
\label{comptable}
\centering
\begin{tabular}{c|c|c|c}
\hline
\bfseries Opening  & \bfseries Computation Time& \bfseries Max Velocity & \bfseries Success Rate\\
\bfseries Rotation ($^{\circ}$) & (second) & (m/s) & (\%) \\
\hline\hline 
  0 & 17.1 & 0.3 & 100.0\\
 30 & 23.2 & 2.1 & 100.0\\
 50 & 32.5 & 6.4 & 89.47\\
 85 & 52.4 & 15.2 & 70.0\\
\hline
\end{tabular}
\end{table}

\section{Discussion}

Compared to other sampling-based planners, the advantage of KAT is using the dual-direction control scheme to generate local paths around narrow passages, saving a large amount of computation time from sampling in the high dimensional state space. This innovation reduces the time complexity dramatically and can be extended to other similar motion planning problems, where a particular subset of the problem poses a much higher challenge than others. 

Although the KAT algorithm has proven to be able to successfully generate aggressive flying patterns in a simulated environment, it could conceivably encounter some difficulties during implementation in real world environments because the execution of a trajectory will always subject to drift in practice. A robust controller might be required to resolve this issue and make implementation more feasible. Future work on KAT may include generalizing the concept of near-holonomic set on other robotic systems, and designing an evaluation method of the success rate of the local aggressive paths.

\section{Conclusion}

% The value of the KAT algorithm is that by identifying the hardest part of the problem first, it prevents the chance of failing midway while executing the algorithm.

In this work, we have proposed the KAT path planning algorithm for systems with nonholonomic constraints. The KAT is aimed to solve planning problem where aggressive maneuver is required to pass the narrow passages. The algorithm reduces the computation cost significantly by first identifying the states allow the robot to pass the narrow passages and then planning the local path exploiting the dual-direction control scheme. In the simulation, KAT can efficiently plan a quadcopter through two walls with tilted holes, showing it is a effective planner for aggressive trajectories.

%\section*{Acknowledgment}
%The authors would like to thank Prof.Berenson for inspirational discussions. 

% \ifCLASSOPTIONcaptionsoff
  \newpage
% \fi

\bibliographystyle{ieeetr}
\bibliography{./bibtex/bib/598}

\end{document}